\documentstyle[aps,prl,multicol,psfig]{revtex}

\begin{document}
\draft
\title{Weak anisotropy  and disorder dependence of the in-plane  magnetoresistance
 in high mobility (100) Si-inversion layers}

\author{V.\ M.\ Pudalov$^a$
G.\ Brunthaler$^b$, A.\ Prinz$^b$, and G.\ Bauer$^b$}
\address{$^a$ P.\ N.\ Lebedev Physics
Institute, 119991 Moscow, Russia.}
\address{$^b$ Institut f\"{u}r Halbleiterphysik,
Johannes Kepler Universt\"{a}t, Linz, Austria}
\date{\today}
\maketitle

\begin{abstract}
We report studies of the magnetoresistance (MR) in a
two-dimensional  electron system in (100) Si-inversion layers, for
perpendicular and parallel orientations of the current with
respect to the magnetic field in the 2D-plane. The
magnetoresistance is almost isotropic; this result does not
support the suggestion  of the orbital origin of the MR in Si-inversion layer.
In the hopping regime, however, the MR
contains a weak  anisotropic component that is non-monotonic in
magnetic field. We found that the field, at which
the MR saturates, for different samples varies by a factor of two,
being lower or higher than the field of complete spin polarization of free carriers.
Therefore, the saturation of the MR can not be identified with
the spin polarization of free carriers.

\end{abstract}

\begin{multicols}{2}

For low carrier densities, in the vicinity of what is known as the
`metal-insulator transition  (MIT) in two dimensions' \cite{rmp}, a magnetic
field $B_{\parallel}$ applied in the plane of the 2D carriers
suppresses the anomalous metallic conduction
\cite{PRL97,bparallel_97,simmons_pGaAs,okamoto,yoon99}. The
resistivity $\rho$ measured at low temperatures ($T\ll E_F$) raises
roughly proportionally to $B_{\parallel}^2$ and further saturates
\cite{PRL97,bparallel_97} (Fermi energy,
$E_F$, through the paper is in units of K). The observed strong
magnetoresistance (MR) and subsequent saturation of the
resistivity in high fields is often interpreted as a manifestation
of the spin alignment of free carriers by the parallel field
\cite{okamoto}. Recently, in Ref.~\cite{das_parallel}, the strong
MR was suggested to be an orbital effect  caused by the transition
from 2D to 3D behavior, when the magnetic length $a_B$ becomes
smaller than the thickness of the quasi-2D layer. We note however,
that there is a third possibility in which the parallel field also
acts  on the spins of the localized carriers (or of the
donor/acceptor bound states), and causes a corresponding increase
in Coulomb scattering. Such a possibility is particularly inherent
in models \cite{am,klap_das,kozub} which consider interface localized
states.

In order to determine  how universal is the  magnetoresistance in parallel field,
we conducted  systematic measurements on a variety of samples  with
different mobilities.
Despite the qualitative
similarity of the magnetoresistance
in all samples,
the characteristic magnetic field at which MR saturates, $B_{\rm sat}$, is found
to be non-universal.
$B_{\rm sat}$ varies  by
a factor of 2 from the highest to the lowest mobility sample
being, respectively,  either lower or higher
than  the  field of complete spin polarization.
This result demonstrates that the saturation of the
magnetoresistance can not be identified with a field of complete
spin polarization of the mobile electrons in Si-MOS samples, the
parameter which should depend only on the carrier density
 \cite{gm} but not on a sample.

To  probe the relevance of the spin and orbital effects, we have
performed measurements of the MR for a magnetic field in the 2D
plane, with the bias current $j$ directed perpendicular and parallel to the
field. In the explored range of densities on both sides of the
MIT, and in magnetic fields   $(0 \div 12)$\,T, we found {\em no
strong anisotropy} in the MR.
This does not support the idea of the orbital origin of the MR in Si-MOS
structures,
but might be consistent with  spin-related mechanisms.
We observed only a  weak anisotropy in the MR, $\Delta
\rho= (\rho_{\parallel}
 - \rho_{\perp})/2\overline{\rho} \sim 5\%$, which in the hopping regime
exhibits a
non-monotonic
dependence on the magnetic
field. To the best of our knowledge, such an anisotropy in the (100)
crystal plane  has not been observed earlier in Si-MOS systems (a
monotonic anisotropy of the MR was reported in Ref.
\cite{papadakis_aniso99} for the anisotropic GaAs/AlGaAs (311)
plane).
The anisotropic
non-monotonic component of the MR indicates an effect of the
spin-orbit (SO) coupling on the electron transport in the vicinity
of the MIT.

\begin{figure}
\vspace{0.1in}
\centerline{
\psfig{figure=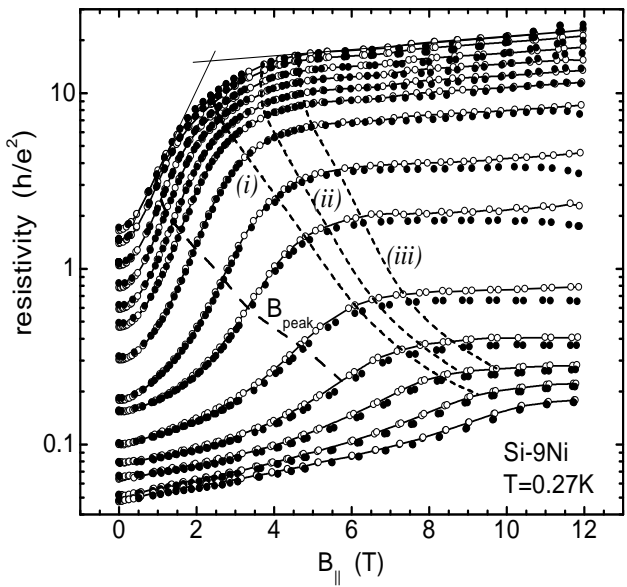,width=235pt,height=220pt}
}
\begin{minipage}{3.2in}
\vspace{0.1in}
\caption{Resistivity vs in-plane magnetic field. Open symbols are
for $j \parallel B$, closed symbols for $j \perp B$. Densities $n$
in units of $10^{11}$\,cm$^{-2}$ are (from top to bottom): 0.748,
0.759, 0.78, 0.80, 0.835, 0.858, 0.913, 1.023, 1.133, 1.353,
1.573, 1.793, 2.013, $2.233\times 10^{11}$cm$^{-2}$. Dashed curves
{\t i, ii, iii} correspond to three definitions of the ``saturation
field'' as described in the text, curve $B_{\rm peak}$ depicts the
position of the anisotropy peak. Two tangents at the most top
curve illustrate the definition (i) of  $B_{\rm sat}$.}
\label{fig1}
\end{minipage}
\end{figure}

The ac-measurements (3\,Hz) of the resistivity  were performed at
temperatures $0.27 - 0.3$\,K on five (100)~Si-MOS samples:  Si-9Nj
(peak mobility $\mu^{peak}=4.3$m$^2$/Vs at $T=0.3$\,K), Si-153
(3.8\,m$^2$/Vs),  Si-15a (3.2\,m$^2$/Vs), Si22a
(2.7\,m$^2$/Vs), and Si43a (1.96\,m$^2$/Vs). The samples were
lithographically defined as rectangular Hall bars of the size
$0.8 \times 5$\,mm (first four samples) and $0.256\times 2.5$\,mm
(the last one);
their long sides (current direction) were aligned along [010].
All samples exhibited qualitatively similar behavior of
$R(B_{\parallel})$. Typical field dependent traces of the
resistivity are shown in Fig. \ref{fig1}. Similar to that
reported earlier \cite{PRL97,bparallel_97}, resistivity grows
with magnetic field and then saturates above a certain field,
$B_{\rm sat}$. To obtain the data for different orientations of
$j$  relative to $B_{\parallel}$, the sample was rotated {\it in
situ} at low temperature. The carrier density $n$ was varied by the gate voltage
and its value was  determined from the
Shubnikov-de Haas (SdH) oscillations,
quantum Hall effect \cite{QHE/I}, and Hall voltage.
For the studied samples, even through the MIT \cite{QHE/I,Halldensity},
the density did not deviate more than by 7\% from
the value extrapolated from the high density regime \cite{density}.

Figure 1 demonstrates the absence of a large anisotropy  in
the MR, and the presence of a minor anisotropy ($\lesssim 10\%$)
in high fields and for intermediate densities with
$\rho(j\parallel B)$ systematically exceeding  $\rho(j\perp B)$
\cite{shift_in_n}. Both these results disagree with the
predictions of Ref. \cite{das_parallel} where the MR was
associated with a crossover from 2D to 3D behavior with increasing
parallel field.
In order to quantify the saturation field $B_{\rm sat}$, we used
three different empiric definitions for $B_{\rm sat}$ as a field
corresponding to the interception of the tangents (as
demonstrated in Fig.~1) on the logarithmic  (i) or  linear (ii)
scale of resistance, and  (iii) corresponding to an increase of
the MR to 97.5\% of its maximum value. Three  dashed lines in
Fig.~1 connect the $\rho(B_{\rm sat})$-points determined
according to the definitions  (i) to (iii).

The $B_{\rm sat}$-data according to the
definition (i) are plotted in Fig.~2 vs electron density.
In accord with other results
\cite{bparallel_97,okamoto,yoon99,vitkalov,vitkalov_SdH}, we find
that $B_{\rm sat}$
increases approximately linearly with the density.  We therefore
assume first that the saturation field  corresponds to the complete
polarization of the 2D electron system,
\begin{equation}
B_{\rm pol}  = \frac{2 E_F}{g^*\mu_B}= \frac{n}{g^*m^*}\frac{h}{e},
\label{eq1}
\end{equation}
where $g^*$ is  the effective Land\'{e} $g-$factor,
$m^*$ the effective mass and $\mu_B$ is the Bohr
magneton (valley degeneracy $g_v =2$ for (100)Si crystal plane).
In Ref.~\cite{gm}, the $g^*m^*$ values   were
measured
as a function of the carrier density; it was also
found that $g^*m^*$ is sample independent. Therefore,
$B_{\rm pol}$ must  also be a universal function of the carrier density.
Contrary to this expectation, we found $B_{\rm sat}(n)$
 to be noticeably different for
different samples  (see Fig.~2).
This result, evidently, does not support the proposed model
\cite{dolgopolov2000} where the MR is related to the
screening radius (or to  the density of states at $E_F$)
which depends on the free carrier density
only.

\begin{figure}
\centerline{
\psfig{figure=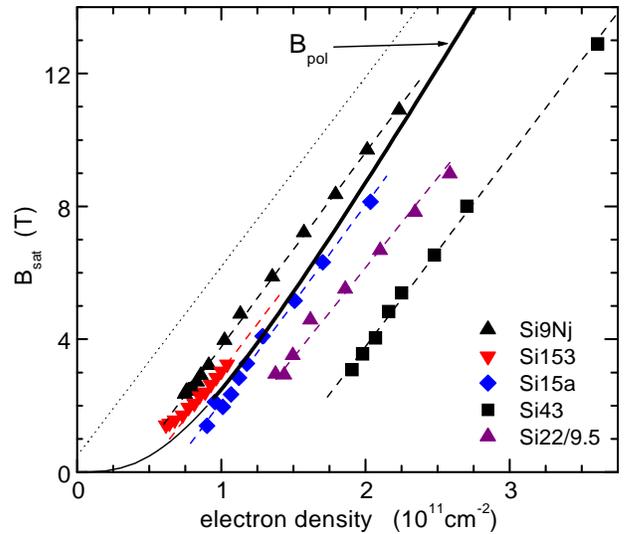,width=230pt}
}
\begin{minipage}{3.2in}
\vspace{0.1in}
\caption{$B_{\rm sat}$ vs electron density for five samples.
Dashed lines are the linear
fits. Continuous line $B_{\rm pol}(n)$ shows the
calculated field of the spin  polarization, plotted over the  range of densities explored
in Ref.~\protect\cite{gm} (thick line),
and extended to lower densities (thin line). Dotted line shows an empiric
$B_{\rm sat}(n)$ dependence  with the offset
extrapolated to $1/\mu^{\rm peak} =0$, as discussed in the text.}
\label{fig2}
\end{minipage}
\end{figure}

The  $B_{\rm sat}$ values depend on the choice of the
 empiric definition, however,
 the difference between different samples persists within any  definition of this parameter.
We conclude therefore
that the magnetic field value, at which the
magnetoresistance saturates, {\em does not reflect
the alignment of the free carrier spins}.
In Fig. 2, for comparison, we plotted also the field $B_{\rm pol}$
corresponding to complete spin
polarization which was calculated according to Eq.~(1)
and using the $g^*m^*(n)$ values
determined in Ref.~\cite{gm}.
Clearly, for different samples, the magnetoresistance
saturates at fields, which could be either lower or higher
than the field of complete spin-polarization, $B_{\rm pol}$.
For some samples and over a limited density range,
$B_{\rm sat}$ is rather close to $B_{\rm pol}$,
as was noted in Ref.~\cite{vitkalov,vitkalov_SdH}.

The $B_{\rm sat}(n)$ dependences shown in Fig.~2
have similar slopes
($dB_{\rm sat}/dn \sim 5.7$\,T$/10^{11}\mbox{cm}^{-2}$), but
each of them  extrapolates to zero at a
sample-dependent finite offset density, $n_{\rm d}$. The latter
 is of the order of the `critical density' $n_c$ for
the MIT at zero field. More specifically, $n_{\rm d}$
varies  from approximately $0.4n_c$ for the highest mobility
sample, Si9Nj, to $\approx n_c$ for the lowest mobility one, Si-43a
\cite{note_cooldown}.
The offset $n_{\rm d}$  and the inverse peak mobility (a measure of the sample disorder)
are in an apparent linear relation (values of $\mu^{peak}$ for each sample are given above).
The dotted curve in Fig.~2 represents  a schematic dependence $B_{\rm sat}(n)$
with the same slope as the experimental data and with the offset
extrapolated linearly to the infinite mobility.

\begin{figure}
\centerline{
\psfig{figure=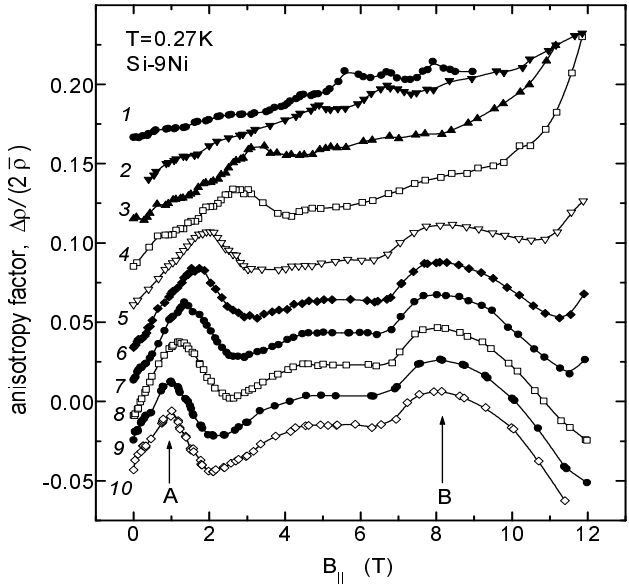,width=235pt,height=220pt}
}
\begin{minipage}{3.2in}
\vspace{0.1in}
\caption{Anisotropy of the MR vs parallel magnetic field.
For clarity, the curves are shifted vertically by
0.02 relative to each other. Arrows `A' and `B'
show two major features.
The densities $n$ in units of
$10^{11}$\,cm$^{-2}$ are (from  bottom to top):
0.748,  0.759, 0.78, 0.80, 0.835, 0.858, 0.913,
1.023, 1.133, 1.353.}
\label{fig3}
\end{minipage}
\end{figure}

It was suggested in Ref.~\cite{vitkalov} that the non-zero value of $n_{\rm d}$ is
caused by a spontaneous spin polarization of the electron system.
Clearly, such an assumption would lead to a paradox, since more and more
disordered samples would exhibit this phenomenon at higher and higher
density (or lower inter-electron interaction strength).
On the other hand,  no disorder- (or sample-) dependence is found
in the spin susceptibility ($\chi\propto g^*m^*$)
directly measured in Ref.~\cite{gm}
down to $n=1\times 10^{11}$\,cm$^{-2}$ (i.e.,  to
the critical density of the MIT). It is therefore rather likely that
the offset value, $n_{\rm d}$, and the whole effect of the magnetoresistance
saturation in Si-MOS samples are related mainly with a
physics of the disorder or of the bound states,
rather than with intrinsic  properties of
free electrons in the 2D system.

The
non-universality of $B_{\rm sat}$ reflects
a breakdown of the models which assume the magnetoresistance to be related only with
 the Zeeman energy of free carriers. On the other hand, such a behavior is anticipated for
 the models which consider the magnetoresistance to be a result of
 floating up of the band of localized states  e.g., such as considered
in Refs.~\cite{kozub,disorder}.

We now turn to  a comparison  of the magnetoresistance measured with different
current direction; a more accurate analysis
reveals their weak non-monotonic difference,  $\Delta \rho = \rho (j \parallel B) - \rho
(j \perp B)$. Arrows in Fig.~3 mark two major features,  a
density-dependent peak `A' and a density-independent
broad maximum `B'. Both peaks are well pronounced only in the hopping
regime (i.e., for $\rho(B,n) \gtrsim (0.2-2)h/e^2$,
depending on the magnetic field \cite{disorder}) and disappear
for densities  $n \gtrsim 1\times 10^{11}$cm$^{-2}$.
Whereas the monotonic component of the MR anisotropy (Fig.~3)
might in principle be related to the effects of finite
thickness of the 2D layer \cite{das_parallel}, it is not the case
for the peak structure of the MR. We also note that both peaks
can not be caused by a perpendicular field component due to a
minor misalignment of the sample plane  \cite{tilt}.

\begin{figure}
\centerline{
\psfig{figure=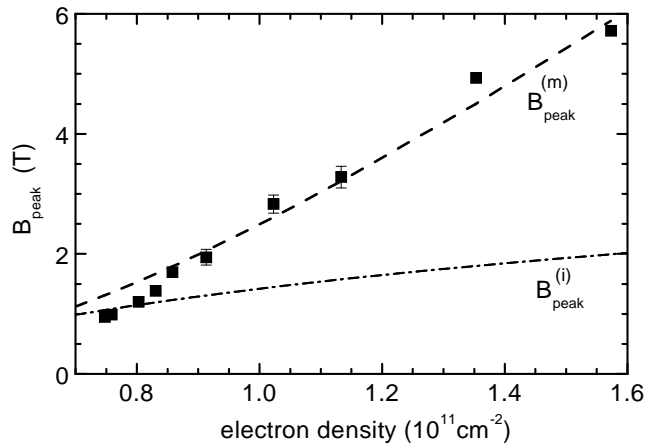,width=240pt}
}
\begin{minipage}{3.2in}
\caption{Magnetic field position of the peak `A'
in the anisotropic component of the MR  vs electron
density. Solid symbols are the measured data, dashed curve
 represents the theoretical value $B_{\rm peak}^{(m)}$
 \protect\cite{raimondi}, dash-dotted curve is for $B_{\rm peak}^{(i)}$
 \protect\cite{raikh_aniso}. Both theoretical curves are calculated using $g^*m^*(n)$
data measured in Ref. \protect\cite{gm}.}
\label{fig4}
\end{minipage}
\end{figure}
\vspace{0.1in}

At present,
there is no explanation for the origin of the
density-independent peak `B'. As for the peak `A',
its characteristic magnetic field  $B_{\rm peak}$
increases
with density as Fig.~4 shows.
A  peak of a similar shape in the MR anisotropy was theoretically
predicted to occur, due to the interplay between the spin-orbit
and Zeeman coupling \cite{raikh_aniso,raimondi}.
In particular,  Chen et al. \cite{raikh_aniso} considered
the hopping (insulating) conduction regime and found that the
peak in the anisotropy should take place  at a field
$B_{\rm peak}^{(i)} = \alpha k_F/g^*\mu_B$
(where $\alpha$ is the spin-orbit coupling constant,
and $k_F$ is the Fermi momentum). Raimondi et al. \cite{raimondi} considered
the diffusive (metallic) transport regime
and found that a peak in anisotropy occurs at a field
$B_{\rm peak}^{(m)}= 2E_F/g^*\mu_B$.
For comparison, we plotted in Fig.~4  both  theoretical dependences $B_{\rm peak}(n)$.
In the calculations we did not use  adjustable parameters \cite{note_alpha};
for the renormalized $g^*$ and $g^*m^*$ values we used the
experimentally determined data \cite{gm}.

There is a similarity between the measured location of the peak `A' and   $B_{\rm peak}^{(m)}$
calculated in Ref. \cite{raimondi}, though the calculations are done
for the diffusive transport regime
whereas in the experiment at such high fields, the transport is hopping \cite{disorder}.
The inconsistency with the $B_{\rm peak}^{(i)}$ calculated in Ref.~\cite{raikh_aniso}
for the hopping transport
cannot be eliminated by selecting any   value of $\alpha$, the parameter
which is supposed to be density independent.

In conclusion, we have shown that  at low carrier densities in
the vicinity of the MIT,
the parallel-field
magnetoresistance in (100) Si-inversion layers, is almost independent of the
relative orientation of the bias current and magnetic field. This is
inconsistent with the orbital origin of the strong MR and supports
its spin origin.
We have observed  a
weak ($\approx 5\%$) anisotropy of the
MR in the hopping regime, which is non-monotonic as a function of the magnetic field.
In particular, it exhibits a density-dependent sharp peak and
a broad maximum. We compared  the sharp peak in the anisotropy of the magnetoresistance
with the one theoretically predicted and related with
the interplay between
the  spin-orbit and Zeeman coupling. We found
the peak shape and the magnetic field position to be
consistent with the theoretical predictions.
We found that the field at which magnetoresistance of Si-inversion layers saturates  is a sample-
(and, apparently, a disorder-)
dependent parameter. For a variety of samples studied,
the `saturation' field
may be substantially larger or smaller than
the field of complete spin polarization. Therefore, the saturation of the
MR can not be identified with the complete spin-alignment of free
carriers.

Authors acknowledge discussions with B.\ L.\ Altshuler, D.\ L.\ Maslov,
M.\ E.\ Gershenson,  H.\ Kojima, and M.\ E.\ Raikh.
The work was supported by FWF (project No 13439), INTAS, NATO,
NSF, and the Russian programs RFBR, \lq Physics of solid state
nanostructures', \lq Statistical physics', \lq Integration' and
`The State support of the leading scientific schools'.

\end{multicols}


\begin{references}
\bibitem{density}Measurements of Shubnikov-de Haas
oscillations in parallel magnetic field  \protect\cite{gm}, on similar samples,
have shown that the changes in density are $\leq 5\%$, up to the fields corresponding to
70\% of the spin polarization.
\bibitem{shift_in_n} A minor difference in $\rho$
taken with
$j_{\perp}B$ and $j_{\parallel}B$ can be also seen at $B=0$. This
indicates a minor difference in the carrier density, $\sim 2.5\times
10^{8}$cm$^{-2}$ in these two measurement runs.
\bibitem{note_cooldown}The samples were cooled down in the same way
so that the potential well was of the same shape.
\bibitem{tilt}A perpendicular field component
due to its possible misalignment by the angle $\theta \leq 2^\circ$,
$B_{\perp} = \sin(\theta) B_{\rm total}$,
in the vicinity of the MIT should induce the re-entrant
insulating state with peaks in $\rho_{xx}$ at filling factors
$\nu =1.5$ and $2.5 - 3$.
The peaks should thus be located at $B_{\rm total}= n/[0.242 \nu
\sin(\theta)]$, where $B$ is in T and $n$ is in
$10^{11}$cm$^{-2}$.
The magnetic field for the observed peak `B' is however density independent, whereas
the  $dB_{\rm peak}/dn$ for the peak `A' is about $10\times$
smaller than the above estimate; an assumption of misalignment by $20^{\circ}$ is unrealistic.
\bibitem{note_alpha} For $\alpha$ we took a value $6\times 10^{-6}$K\,cm
\protect \cite{pudalov_beats} estimated from a beating
period of the  Shubnikov-de Haas oscillations (though for higher
densities).
\end{references}
\end{document}